# Monolithically integrated waveguide-coupled single-frequency microlaser on erbium-doped thin film lithium niobate

YOUTING LIANG,[1,2] JUNXIA ZHOU,[1,2] RONGBO WU,[3,4,8] ZHIWEI FANG,[2,9] ZHAOXIANG LIU,[2] SHUPENG YU,[3,4] DIFENG YIN,[3,4] HAISU ZHANG,[2] YUAN ZHOU,[3,4] JIAN LIU,[2] ZHENHUA WANG,[2] MIN WANG,[2] AND YA CHENG[1,2,3,5,6,7,10]

[1]*State Key Laboratory of Precision Spectroscopy, East China Normal University, Shanghai 200062, China*
[2]*The Extreme Optoelectromechanics Laboratory (XXL), School of Physics and Electronic Science, East China Normal University, Shanghai 200241, China*
[3]*State Key Laboratory of High Field Laser Physics and CAS Center for Excellence in Ultra-intense Laser Science, Shanghai Institute of Optics and Fine Mechanics (SIOM), Chinese Academy of Sciences (CAS), Shanghai 201800, China*
[4]*Center of Materials Science and Optoelectronics Engineering, University of Chinese Academy of Sciences, Beijing 100049, China*
[5]*Collaborative Innovation Center of Extreme Optics, Shanxi University, Taiyuan 030006, China.*
[6]*Collaborative Innovation Center of Light Manipulations and Applications, Shandong Normal University, Jinan 250358, People's Republic of China*
[7]*Shanghai Research Center for Quantum Sciences, Shanghai 201315, China*
[8] rbwu@siom.ac.cn
[9] zwfang@phy.ecnu.edu.cn
[10] ya.cheng@siom.ac.cn

**We overcome the difficulty in realizing a monolithic waveguide-coupled microring laser integrated on erbium-doped thin film lithium niobate (Er: TFLN) using photolithography assisted chemo-mechanical etching (PLACE) technique. We demonstrate an integrated single-frequency microring laser operating around 1531 nm wavelength. The PLACE technique, enabling integrated Er: TFLN photonics with low propagation loss, can thus be used to realize low cost mass production of monolithic on-chip microlasers with applications ranging from optical communication and photonic integrated circuit (PIC) to precision metrology and large-scale sensing.**

Featured with its broad optical transparency, high refractive index, large electro-optical and nonlinear optical coefficients, the integrated thin film lithium niobate (TFLN) photonics have emerged as a promising platform for realization of high-performance photonic integrated circuit (PIC) both for classical and quantum applications [1-7]. Moreover, since the lithium niobate (LN) crystal proves to be an attractive host material for rare earth ions (REIs), various on-chip microlasers and waveguide amplifiers have been successfully fabricated on the REI-doped TFLN recently [8-17]. The on-chip microlasers are first demonstrated with multi-frequency lasing. Single-frequency microlaser is further realized through the Vernier effect with two evanescently coupled microresonators, though the delicate resonant mode matching between the coupled microresonators imposes stringent requirements on the fabrication precision [18-21]. Recently, the single-frequency output in a single Er: TFLN microring resonator by introducing mode-dependent loss and gain competition has been reported [22]. However, such on-chip single-frequency microlaser was fabricated by electron beam lithography (EBL) with limited exposure areas. Here, we demonstrate an integrated single-frequency microring laser operating around 1531 nm wavelength fabricated by the photolithography assisted chemo-mechanical etching (PLACE) technique. We overcome the difficulty in realizing monolithic integration of waveguide-coupled microring laser by the PLACE technique [23, 24]. Benefited from the large writing field and high writing speed of PLACE, in sharp contrast with the rather limited writing field of EBL, the PLACE technique provide superior agility and scalability concerning the footprint and power capacity of the on-chip microlaser based on the microring resonators, promising in low cost on-chip microlasers, with broad applications ranging from optical communication and photonic integrated circuit (PIC) to precision metrology and large-scale sensing.

The microring resonator was fabricated on an Er: TFLN wafer by the PLACE technique. The Er: TFLN wafer was prepared by bonding a Z-cut $Er^{3+}$-doped 500-nm-thick TFLN onto a holder wafer with 2-μm-thick SiO2 and 500-μm-thick silicon. The concentration of Er³⁺ ions in the TFLN is 1 mol%. The scanning electron microscope (SEM) image of the cross section of the designed waveguide-ring coupling region is shown in Figure 1(a), and the tilt angle of the SEM image is about 52°. The Er³⁺-doped TFLN is coated with Au thin film for the sake of focused ion beam (FIB) cutting and clear imaging in the SEM. The Er³⁺-doped TFLN is shown in purple and the silica layer is shown in gray. The gap between the straight waveguide and the ring resonator was set to be 4.8 um. Er³⁺-doped TFLN waveguides with a top width of 935 nm and a thickness of 465 nm are located on the top of a 2-μm-thick SiO2 layer. The fabricated LN waveguide is slightly thinner than the original thickness of the TFLN due to the second chemo-mechanical polishing (CMP) process. Besides, due to the anisotropic polishing rate of the CMP process, the thickness at the bottom of the waveguide-ring gap is 300 nm while the thickness of the outer edge of waveguide is 230 nm. The simulated electric field profiles of the symmetry and anti-symmetry mode of the coupling region are shown in Figures 1(b) and (c). The coupling efficiency $K$ in the coupling area can be estimated as 0.04 dB at 1550 nm using equation $K = (1 - \cos((2\pi \Delta n_L)/\lambda))/2$, where $L$ = 50 um is the coupling length, $n_1$ and $n_2$ are the effective refractive index of the symmetry and anti-symmetry mode, respectively, The required coupling efficiency $K$ for critical coupling between the waveguide and the microring is calculated as 0.03 dB using $K = \alpha L_r$, where $\alpha \approx 0.3$ dB/cm is the waveguide propagation loss, $L_r$ = 1 mm is the perimeter of the microring resonator.

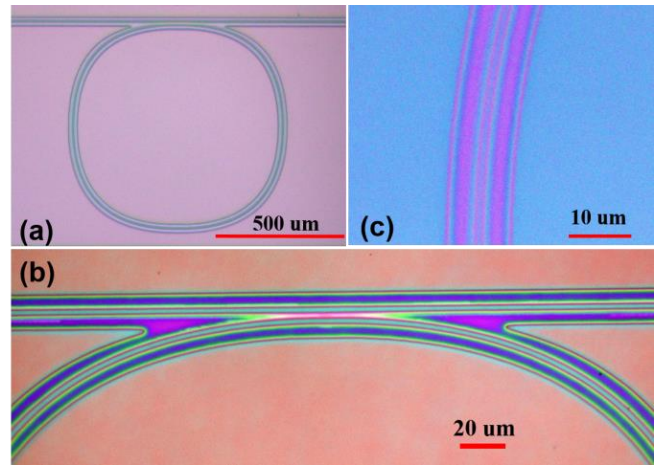

Fig. 2. (a) Optical micrograph of the fabricated optical devices. (b) Close-up optical micrograph of the LN waveguide. (c) Close-up optical micrograph of the coupling region of the microring resonator.

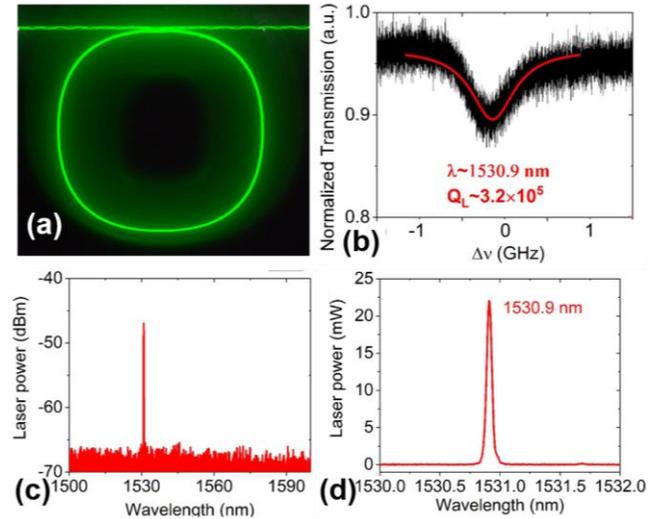

Fig. 3 (a) the green upconversion fluorescence of Er: TFLN microring when the 976 nm pump diode laser is injected. (b) A transmission spectrum of the microring resonator with a loaded Q factor of $3.2 \times 10^5$ at the wavelength 1530.9 nm. (c) Output spectrum of the Er: TFLN microring laser shows only one lasing emission peak. (d) The enlarged spectrum around wavelength 1530.9 nm

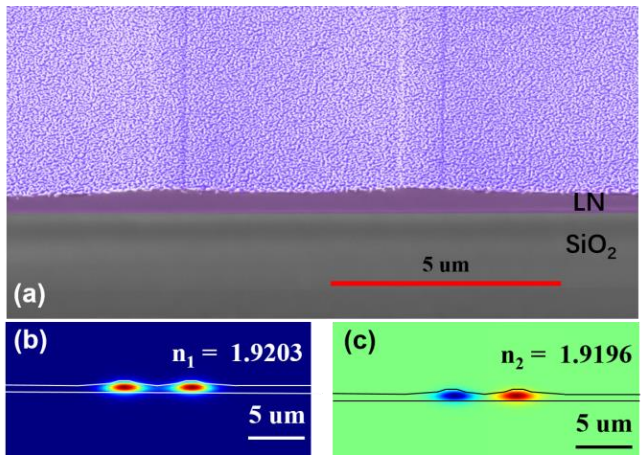

Fig. 1. (a)The false color scanning electron microscope (SEM) images of the cross section of the gap between waveguide and microring, the Er³⁺-doped TFLN is shown in purple and the silica layer is shown in gray, the Au thin film is coated on the surface to benefit focused ion beam cutting and clear imaging the cross section of the coupling area in the SEM. The simulated electric field profile of the (b) symmetry and (c) anti-symmetry mode.

The fabricated microring laser is shown in Figure 2(a), the Er³⁺-doped LN microring resonator is design with improved quality factors using quarter Bezier curves due to mode match in coupling region and lower bending loss than the conventional circular ring resonator [25]. The length of the perimeter is about 1 mm.. Figure 2(b) shows the close-up optical micrograph of the LN waveguide. The gap in both sides of LN waveguide is about 4 um. Figure 2(c) shows the close-up optical micrograph of the coupling region of the microring resonator where the mode match between the straight waveguide and the microring resonator with quarter Bezier curves. The different thicknesses of the gap bottom and the waveguide outer edge are clearly visible by the color changes in the optical micrograph of the coupling region.

Light from a diode laser (CM97-1000-76PM, II-VI Laser Inc.) and a tunable telecom laser (CTL 1550, TOPTICA Photonics Inc.) are coupled into and collected from the waveguide facets using lensed fibers. The polarization of the pump light was adjusted by a 3-paddle fiber polarization controller (FPC561, Thorlabs Inc.). The power of the input pump laser was monitored by a power meter (PM100D, Thorlabs Inc.). A photodetector (New focus 1811, Newport Inc.) was placed in the fiber path for the quality (Q) factor measurements of resonant modes in the microring resonator. The spectra of the output beam were measured by an optical spectrum analyzer (OSA: AQ6375B, YOKOGAWA Inc.). Figure 3(a) shows the green upconversion fluorescence micrograph when the 976 nm diode laser is injected into the Er: TFLN microring. Figure 3(b) shows a transmission spectrum of a microring resonator with a loaded Q factor of $3.2 \times 10^5$ at the wavelength 1530.9 nm. As shown in Figure 3(c), only one lasing emission peak is collected by OSA in the wavelength range of 1500 nm -1600 nm probably due to mode-dependent loss and gain competition. Figure 3(d) shows the enlarged spectrum around lasing emission, featuring a linewidth of 50 pm at 1530.9 nm which is limited by the resolution of the OSA (~0.01 nm).

Figure 4(a) shows the spectra of the Er: TFLN microring laser at the increasing pump powers. The dependence of the lasing power of the Er: TFLN microring laser on the injected pump power is illustrated in Figure 4(b). The lasing threshold is found to be around 24.5 mW by linear fitting. The derived threshold is higher than previous work due to the longer perimeter of the $Er^{3+}$-doped LN microring which necessitates higher pump power to invert erbium ions [18-20].

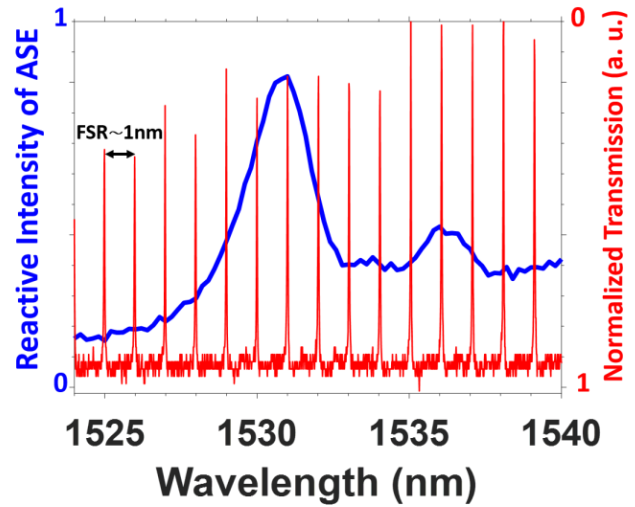

Fig. 5. ASE of an Er: TFLN straight waveguide (blue) and normalized transmission of Er: TFLN microring.

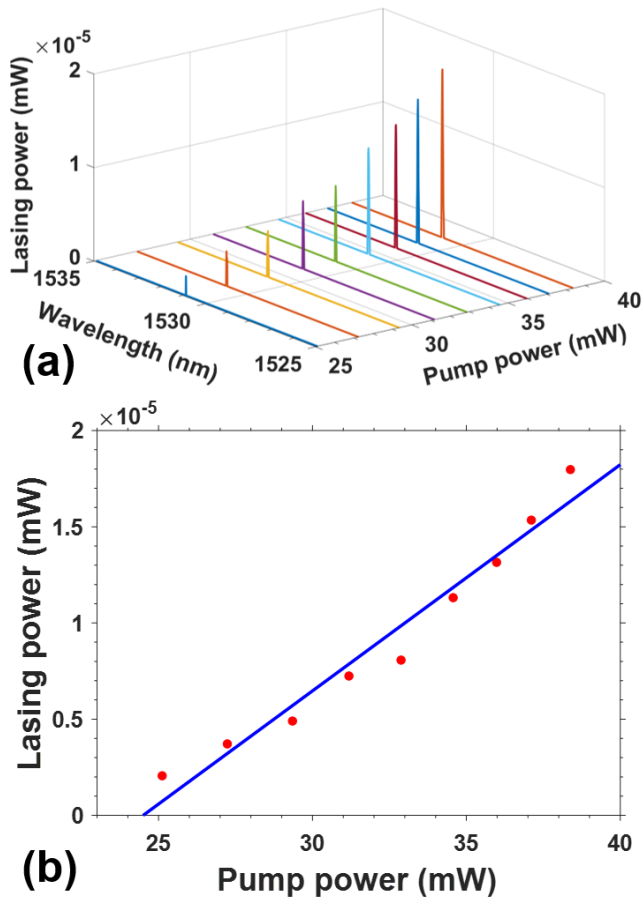

Fig.4. (a) The spectra of the Er: TFLN microring laser at the increasing pump powers. (b) The dependence of the lasing power of the Er: TFLN microring laser on the input pump power.

As shown in Figure 5, we measured both the transmission spectrum of the Er:TFLN microring (shown in red) and the amplified spontaneous emission(ASE) spectrum (shown in blue) from a Er:TFLN waveguide. The free spectral range (FSR) of Er:TFLN microring is about 1 nm. The ASE of Er:TFLN waveguide has a sharp gain peak near 1531 nm. It can be seen that there is one resonance of the microring which is coincident with the sharp gain peak of ASE near 1531 nm. Furthermore, the identical Q factors for all the resonances shown in Figure 5 indicate comparable losses of the microring in the investigated spectral range. Therefore, the resonance around the gain peak will laser first and then grow faster when increasing the injected pump powers, which in turn can suppress other resonances from lasing due to gain competition. Further systematic investigation is needed to clarify the observed single-frequency lasing behavior of the waveguide-coupled microring resonator.

In conclusion, we demonstrate an integrated single-frequency microring laser operating around 1531 nm wavelength fabricated by photolithography assisted chemo-mechanical etching technique. The PLACE technique can achieve the further limits the maximum output power in on-chip microlaser based on microring resonators. Moreover, the PLACE technique, in conjunction with integrated low propagation loss lithium niobate photonics, can be used to realize low cost and mass production of tunable on-chip microlasers, with applications ranging from optical communication and PIC to precision measurement.

**REFERENCES**

1. A. Honardoost, K. Abdelsalam, S.Fathpour, "Rejuvenating a Versatile Photonic Material: Thin-Film Lithium Niobate," Laser Photon. Rev. **14**, 2000088(2020,).
2. J. Lin, F. Bo, Y. Cheng, and J. Xu, "Advances in on-chip photonic devices based on lithium niobate on insulator," Photon. Res. **8**, 1910 (2020).
3. Y. Kong, F. Bo, W. Wang, D. Zheng, H. Liu, G. Zhang, R. Rupp, and J. Xu, "Recent progress in lithium niobate: optical damage, defect simulation, and on‐chip devices" Adv. Mater. **32**, 1806452 (2020).
4. D. Zhu, L. Shao, M.e Yu, R. Cheng, B. Desiatov, C. J. Xin, Y. Hu, J. Holzgrafe, S. Ghosh, A. Shams-Ansari, E. Puma, N. Sinclair, C. Reimer, M. Zhang, and M. Lončar, "Integrated photonics on thin-film lithium niobate," Adv. Opt. Photon. **13**, 242-352 (2021).
5. Y. Jia, L. Wang, and F. Chen, "Ion-cut lithium niobate on insulator technology: Recent advances and perspectives", Appl. Phys. Rev. **8**, 011307 (2021).
6. E. Pelucchi, G. Fagas, I. Aharonovich, D. Englund, E. Figueroa, Q. Gong, H. Hannes, J. Liu, C. -Y. Lu, N. Matsuda, J.-W. Pan, F. Schreck, F. Sciarrino, C. Silberhorn, J. Wang and K. D. Jöns, "The potential and global outlook of integrated photonics for quantum technologies", Nat. Rev. Phys. (2021). https://doi.org/10.1038/s42254-021-00398-z.
7. E. Lomonte, M. A. Wolff, F. Beutel, S. Ferrari, C. Schuck, W. H. P. Pernice, and F. Lenzini, "Single-photon detection and cryogenic reconfigurability in lithium niobate nanophotonic circuits," Nat Commun **12**, 6847 (2021).
8. Z. Wang, Z. Fang, Z. Liu, W. Chu, Y. Zhou, J. Zhang, R. Wu, M. Wang, T. Lu, and Y. Cheng, "On-chip tunable microdisk laser fabricated on $Er^{3+}$-doped lithium niobate on insulator," Opt. Lett. **46**, 380 (2021).

9. Y. A. Liu, X. S. Yan, J. W. Wu, B. Zhu, Y. P. Chen, and X. F. Chen, "On-chip erbium-doped lithium niobate microcavity laser," Sci. China-Phys. Mech. Astron. **64**, 234262 (2021).
10. Q. Luo, Z. Z. Hao, C. Yang, R. Zhang, D. H. Zheng, S. G. Liu, H. D. Liu, F. Bo, Y. F. Kong, G. Q. Zhang, and J. J. Xu, "Microdisk lasers on an erbium doped lithium-niobite chip," Sci. China-Phys. Mech. Astron. **64**, 234263(2021).
11. Q. Luo, C. Yang, R. Zhang, Z. Hao, D. Zheng, H. Liu, X. Yu, F. Gao, F. Bo, Y. Kong, G. Zhang, and J. Xu, "On-chip erbium-doped lithium niobate microring lasers," Opt. Lett. **46**, 3275 (2021).
12. Y. Zhou, Z. Wang, Z. Fang, Z. Liu, H. Zhang, D. Yin, Y. Liang, Z. Zhang, J. Liu, T. Huang, R. Bao, R. Wu, J. Lin, M. Wang, and Y. Cheng, "On-chip microdisk laser on $Yb^{3+}$-doped thin-film lithium niobate," Opt. Lett. **46**, 5651-5654 (2021)
13. J. Zhou, Y. Liang, Z. Liu, W. Chu, H. Zhang, D. Yin, Z. Fang, R. Wu, J. Zhang, W. Chen, Z. Wang, Y. Zhou, M. Wang, and Y. Cheng, "On-chip integrated waveguide amplifiers on Erbium-doped thin film lithium niobate on insulator," Laser Photon. Rev. **15**, 2100030(2021).
14. Z. Chen, Q. Xu, K. Zhang, W.-H. Wong, D.-L. Zhang, E. Y.-B. Pun, and C. Wang, "Efficient erbium-doped thin-film lithium niobate waveguide amplifiers," Opt. Lett. **46**, 1161 (2021).
15. Q. Luo, C. Yang, Z. Hao, R. Zhang, D. Zheng, F. Bo, Y. Kong, G. Zhang, J. Xu, "On-chip erbium-doped lithium niobate waveguide amplifiers," Chin. Opt. Lett. **19**, 060008(2021).
16. M. Cai, K. Wu, J. Xiang, Z. Xiao, T. Li, Ch. Li, and J. Chen, "Erbium-doped lithium niobate thin film waveguide amplifier with 16 dB internal net gain," IEEE J. Select. Topics Quantum Electron. **28**, 1 (2022).
17. Y. Liang, J. Zhou, Z. Liu, H. Zhang, Z. Fang, Y. Zhou, D. Yin, J. Lin, J. Yu, R. Wu, M. Wang, Y. Cheng, "A high-gain cladded waveguide amplifier on erbium doped thin-film lithium niobate fabricated using photolithography assisted chemo-mechanical etching," Nanophotonics (2022). https://doi.org/10.1515/nanoph-2021-0737
18. R. Gao, J. Guan, N. Yao, L. Deng, J. Lin, M. Wang, L. Qiao, Z. Wang, Y. Liang, Y. Zhou, and Y. Cheng, "On-chip ultra-narrow-linewidth single-mode microlaser on lithium niobate on insulator," Opt. Lett. **46**, 3131 (2021).
19. R. Zhang, C. Yang, Z. Hao, D. Jia, Q. Luo, D. Zheng, H. Liu, X. Yu, F. Gao, F. Bo, Y. Kong, G. Zhang and J. Xu, "Integrated lithium niobate single-mode lasers by the Vernier effect," Sci. China Phys. Mech. Astron. **64**, 294216 (2021).
20. X. Liu, X. Yan, Y. a. Liu, H. Li, Y. Chen, and X. Chen, "Tunable single-mode laser on thin film lithium niobate," Opt. Lett. **46**, 5505 (2021)
21. Z. Xiao, K. Wu, M. Cai, T. Li, and J. Chen, "Single-frequency integrated laser on erbium-doped lithium niobate on insulator," Opt. Lett. **46**, 4128 (2021).
22. T. Li, K. Wu, M. Cai, Z. Xiao, H. Zhang, C. Li, J. Xiang, Y. Huang, and J. Chen, "A single-frequency single-resonator laser on erbium-doped lithium niobate on insulator," APL Photonics **6**, 101301 (2021).
23. R. Wu, M. Wang, J. Xu, J. Qi, W. Chu, Z. Fang, J. Zhang, J. Zhou, L. Qiao, Z. Chai, J. Lin and Y. Cheng, "Long Low-Loss-Litium Niobate on Insulator Waveguides with Sub-Nanometer Surface Roughness," Nanomaterials **8**, 910(2018).
24. D. Yin, Y. Zhou, Z. Liu, Z. Wang, H. Zhang, Z. Fang, W. Chu, R. Wu, J. Zhang, W. Chen, M. Wang, and Y. Cheng, "Electro-optically tunable microring laser monolithically integrated on lithium niobate on insulator," Opt. Lett. **46**, 2127 (2021).
25. H. Pishvai Bazargani, J. Azaña, L. Chrostowski, and J. Flueckiger, "Microring resonator design with improved quality factors using quarter Bezier curves," in Conference on Lasers and Electro-Optics: Science and Innovations, (OSA, 2015), paper JTu5A.